\documentclass[12pt]{iopart}

\usepackage{amsfonts}
\usepackage{amscd}

 \font\tensym=msbm10
 \font\sevensym=msbm7
 \font\fivesym=msbm5

 \font\tengoth=eufb10
 \font\sevengoth=eufb7
 \font\fivegoth=eufb5

\newfam\symfam
\textfont\symfam=\tensym
\scriptfont\symfam=\sevensym
\scriptscriptfont\symfam=\fivesym

\newfam\gothfam
\textfont\gothfam=\tengoth
\scriptfont\gothfam=\sevengoth
\scriptscriptfont\gothfam=\fivegoth

\def\hs{\hbox to 3mm{}}
\def\hhs{\hbox to 5cm{}}
\def\ss{\smallskip}

\def\bs{\bigskip}

\def\A{\mathcal{A}}
\def\U{\mathcal{U}}

\def\C{\mathbb{C}}
\def\R{\mathbb{R}}
\def\N{\mathbb{N}}
\def\D{\mathcal{D}}
\def\Diag{{\it Diag}}
\def\LDiag{{\it LDiag}}
\def\DWS{\Delta_{WS}}

\def\HWS{{\cal H}_{WS}}
\def\1H{\mathbf{1}_{\HWS}}
\def\L{\mathbb{L}}
\def\V{\mathbb{V}}

\def\al{\alpha}
\def\be{\beta}
\def\ep{\varepsilon}
\def\la{\lambda}

\def\MQSym{{\bf MQSym}}

\def\lra{\leftrightarrow}

\def\adots{\mathinner{\mkern2mu\raise1pt\hbox{.}
\mkern3mu\raise4pt\hbox{.}\mkern1mu\raise7pt\hbox{.}}}

\def\pointir{\unskip . --- \ignorespaces}

\def\up#1{\raise 1ex\hbox{\footnotesize#1}}

\def\mref#1{{\footnotesize ({\ref{#1}})}}

\newtheorem{expl}{Exemple}[section]

\newtheorem{rem}[expl]{Remark}

\begin{document}

\title[Feynman graphs and related Hopf algebras]
{Feynman graphs and related Hopf algebras}
\author{G H E Duchamp$^{a}$, P Blasiak$^{b}$, A Horzela$^{b}$, K A Penson$^{c}$ 
and A I Solomon$^{c,d}$\vspace{2mm}}

\address{$^a$ Institut Galil\'ee, LIPN, CNRS UMR 7030\linebreak
99 Av. J.-B. Clement, F-93430 Villetaneuse, France\vspace{2mm}}

\address
{$^b$ H. Niewodnicza\'nski Institute of
Nuclear Physics, Polish Academy of Sciences\\
ul. Eliasza-Radzikowskiego 152,  PL 31342 Krak\'ow, Poland\vspace{2mm}}

\address
{$^c$ Laboratoire de Physique Th\'eorique de la Mati\`{e}re Condens\'{e}e\\
Universit\'e Pierre et Marie Curie, CNRS UMR 7600\\
Tour 24 - 2i\`{e}me \'et., 4 pl. Jussieu, F 75252 Paris Cedex 05, France\vspace{2mm}}

\address
{$^e$ The Open University, Physics and Astronomy Department\\
Milton Keynes MK7 6AA, United Kingdom\vspace{2mm}}

\eads{\linebreak\mailto{ ghed@lipn-univ.paris13.fr}, 
\mailto{ pawel.blasiak@ifj.edu.pl},
\mailto{ andrzej.horzela@ifj.edu.pl},
\mailto{penson@lptl.jussieu.fr},
\mailto{ a.i.solomon@open.ac.uk}}


\begin{abstract}
In a recent series of communications  
we have shown that the reordering problem of bosons leads to certain 
combinatorial structures. 
These structures may be associated with a certain graphical description.
In this paper, we show that there is a Hopf Algebra structure associated with this problem which is, in a certain sense, unique.
\end{abstract}

\tableofcontents

\section{Introduction}

In a relatively recent paper Bender, Brody and 
Meister \cite{BBM} introduced a special Field Theory described 
by a product formula (a kind of Hadamard product for 
two exponential generating functions - EGF) in the purpose of proving 
that any sequence of numbers could be described by a suitable set of rules applied to some type of Feynman graphs. 
Inspired by this idea, we have worked out combinatorial consequences of the product and exponential formulas in a recent series 
of papers \cite{GOF2,GOF3,OPM,GOF5,GOF6,GOF7,GOF8}.

Here, we consider two aspects of the product formula for formal power series 
applied to combinatorial field theories. Firstly, we remark that the 
case when the functions involved in the product formula have a constant 
term (equal to one) is of special interest as often these functions give rise to 
substitutional groups. The groups arising from the normal ordering 
problem of boson strings are naturally associated with explicit vector 
fields, or their conjugates, in the case when there is only one 
annihilation operator \cite{OPM,GOF7}. We also consider one-parameter groups of 
operators when several annihilators are present. 
Secondly, we discuss the Feynman-type graph representation resulting 
from the product formula. We show that there is a correspondence 
between the packed integer matrices of the theory of noncommutative 
symmetric functions and the labelled version of these Feynman-type graphs.

We thus obtain a new Hopf algebra structure over the space 
of matrix quasi-symmetric functions that is a natural cocommutative Hopf 
algebra structure on the space of diagrams themselves which originates 
from the formal doubling of variables in the product formula.

\ss
{\sc Aknowledgements} We would like here to express our gratitude to 
Jean-Louis Loday, Jean-Bernard Zuber, Jean-Yves Thibon and Florent Hivert for stimulating interactions on this subject.

\section{Single and double exponentials}

\subsection{One parameter groups and the connected graph theorem}
\subsubsection{Substitutions}

The Weyl algebra $W$ is the $\C$-associative algebra (with 
unit) defined by two generators $a$ and $a^+$ and the unique relation $[a,a^+]=1$. This 
algebra is of Gelfand-Kirillov dimension $2$ and has a basis consisting of the following 
family $\left\{(a^+)^ka^l\right\}_{k,l\geq 0}$.\\
It is known that it is impossible to represent faithfully $a,a^+$ by bounded operators 
in a Banach space, but one often uses the representation 
$a\mapsto \frac{\rm{d}}{\rm{d}x};\ a^+\mapsto x$ as operators acting ``on the line'' or, better said, on the space 
of polynomials $\C[x]$. Through this representation (faithful and coined under the name ``Bargmann-Fock''), 
one sees that we can define a grading on $W$ by the weight function $w(a)=-1;\ w(a^+)=1$.\\
A homogeneous operator (under this grading) $\Omega\in W$ is then of the form
\begin{equation}
\Omega=\sum_{k,l;\ k-l=e}c(k,l)(a^+)^ka^l
\end{equation}
According to whether the excess $e$ is positive or negative, the normal 
ordering of $\Omega^n$ reads

\begin{equation}\label{normal_form1}
\!\!\!\!\!\!\!\!\!\!\!\!\!\!\!\!\!\!{\mathcal N}(\Omega^n)=(a^+)^{ne}\left(\sum_{k=0}^\infty S_\Omega (n,k) (a^+)^ka^k\right)\ \rm{or}\
\left(\sum_{k=0}^\infty S_\Omega (n,k) (a^+)^ka^k\right)(a)^{n|e|}
\end{equation}
We get combinatorial quantities with two indices i.e. an infinite $\N\times\N$ matrix 
$\left\{S_\Omega (n,k)\right\}_{n,k\geq 0}$ which we will call the {\it generalized Stirling matrix of $\Omega$}. 
In fact, it is easily checked that, if the coefficients
 $c(k,l)$ of $\Omega$ are non-negative integers, so are the entries ($S_\Omega (n,k)$) of this matrix.   

\smallskip
Let us give some examples of these generalized Stirling matrices.\\
For $\Omega=a^+a$, one gets the usual matrix of the Stirling numbers of the second kind
\begin{equation}
\left\lceil
{\begin{array}{rrrrrrrr}
1 & 0 & 0 & 0 & 0 & 0 & 0 &\cdots\\
0 & 1 & 0 & 0 & 0 & 0 & 0 &\cdots\\
0 & 1 & 1 & 0 & 0 & 0 & 0 &\cdots\\
0 & 1 & 3 & 1 & 0 & 0 & 0 &\cdots\\
0 & 1 & 7 & 6 & 1 & 0 & 0 &\cdots\\
0 & 1 & 15 & 25 & 10 & 1 & 0 &\cdots\\
0 & 1 & 31 & 90 & 65 & 15 & 1&\cdots\\
\vdots & \vdots & \vdots  & \vdots  & \vdots  & \vdots  & \vdots &\ddots\\
\end{array}}
 \right.
\end{equation}

\smallskip
For $\Omega=a^+aa^+ + a^+$, we obtain
\begin{equation}
\left\lceil
{\begin{array}{rrrrrrrr}
1 	& 0 	& 0 	& 0 	& 0 	& 0 	& 0 	&\cdots\\
2 	& 1 	& 0 	& 0 	& 0 	& 0 	& 0 	&\cdots\\
6 	& 6 	& 1 	& 0 	& 0 	& 0 	& 0 	&\cdots\\
24 	& 36 	& 12 	& 1 	& 0 	& 0 	& 0 	&\cdots\\
120 	&  240 	& 120 	& 20 	& 1 	& 0 	& 0	&\cdots\\
720 	& 1800 & 1200 & 300 	& 30 	& 1 	& 0 	&\cdots\\
5040 & 15120 	&12600 &4200 &630 	& 42 	& 1 	&\cdots\\
\vdots & \vdots & \vdots  & \vdots  & \vdots  & \vdots  & \vdots &\ddots\\
\end{array}}
\right.
\end{equation}
and for $w=a^+aaa^+a^+$
\begin{equation}
\left\lceil
{\begin{array}{rrrrrrrrrr}
1 & 0 & 0 & 0 & 0 & 0 & 0 & 0 & 0 & \cdots\\
2 & 4 & 1 & 0 & 0 & 0 & 0 & 0 & 0 &\cdots\\
12 & 60 & 54 & 14 & 1 & 0 & 0 & 0 & 0 &\cdots\\
144 & 1296 & 2232 & 1296 & 306 & 30 & 1 & 0 & 0 &\cdots\\
2880 & 40320 & 109440 & 105120 & 45000 & 9504 & 1016 & 52 & 1 &\cdots\\
\vdots & \vdots & \vdots  & \vdots  & \vdots  & \vdots  & \vdots & \vdots & \vdots &\ddots\\
\end{array}}
\right.
\end{equation}
Let $\Omega=\sum_{k,l\geq 0}c(k,l)(a^+)^ka^l$ (finite supported sum) be a general term 
of $W$ in normal form and let us call a {\it dominant term}, the sum of monomials with maximum 
length $k+l$. It is not difficult to prove that, if $\Omega$ is homogeneous, the dominant term 
consists of a single monomial $c(k_0,l_0) (a^+)^{k_0}a^{l_0}$. Thus, the dominant term of 
$\Omega^n$ must be $c(k_0,l_0)^n (a^+)^{nk_0}a^{nl_0}$.
Then, for example in the case when $e=k_0-l_0\geq 0$, in the generalized Stirling matrix of $\Omega$, 
the rightmost non-zero coefficient of the line $n$ has address $(n,n.l_0)$ and bears the coefficient $c(k_0,l_0)^n$.    
All these matrices are row-finite and triangular iff $l_0=1$ (which means that no monomial possesses more 
than one $a$).

\begin{rem} i) There is a beautiful combinatorial expression of the normal form of $w^n$ in case 
$w$ is a string in $a$ and $a^+$. The normal form of $w$ is 
\begin{equation}
{\mathcal N}(w)= \sum_{k\geq 0}r(B,k) (a^+)^{r-k}a^{s-k}
\end{equation}
where $r(B,k)$ is the $k$th rook number of a certain board $B$ constructed after $w$
(see  \cite{GOF3,Varvak}, and $r=|w|_{a^+};\ s=|w|_{a}$ 
are the number of occurences of $a^+$ and $a$ in $w$.\\
ii) To each matrix $M\in \C^{\N\times \N}$ of this kind and more generally ``row finite'' matrices 
(which means that, for each $n$, the family $(M(n,k))_{k\in \N}$ is finite supported), one can associate a 
transformation of EGFs (see {\rm \cite{OPM,GOF7}}) $f\mapsto \hat f$ such that, if $f=\sum_{n\geq 0} a_n \frac{z^n}{n!}$ then 
$\hat{f}=\sum_{n\geq 0} b_n \frac{z^n}{n!}$ (with 
$b_n=\sum_{k\geq 0} M(n,k)a_k$).\\
iii) It can be shown that, if no monomial of $\Omega$ possesses more than one $a$, the action of the transformation 
induced by $\Omega$ (through the Bargmann-Fock representation) can be expressed in terms of vector fields or their conjugates, thus 
the one-parameter group $e^{\lambda\Omega}$ acts by substitutions and products {\rm \cite{OPM,GOF7}}.
\end{rem}

\subsubsection{Combinatorial matrices and one-parameter groups}

One can also draw generalized Stirling matrices from another source, namely from 
the combinatorial graph theory.\\
Let $\mathcal{C}$ be a class of graphs such that 
\begin{equation}
\Gamma\in \mathcal{C} {\rm\ iff\ every\ connected\ component\ of\ }\Gamma {\rm\ is\ in\ }\mathcal{C}
\end{equation}
For these classes of graphs, one has the exponential formula \cite{F1,St2,K1} saying roughly that

\begin{equation}
\rm{EGF(all\ graphs)}=e^{\rm{EGF(Connected\ Graphs)}}
\end{equation}
This implies, in particular, that the matrix 
\begin{equation}\!\!\!\!\!\!\!\!\!\!\!\!\!\!\!\!\!\!\!\!\!\!\!\!\!\!\!\!\!\!\!\!\!\!\!\!
M(n,k)={\rm\ number\ of\ graphs\ with\ }n{\rm\ vertices\ and\  
having\ }k{\rm\ connected\ components} 
\end{equation}
is the matrix of a substitution (see \cite{OPM,GOF7}). 
One can prove, using a Zariski-like argument (a polynomial vanishing for every integer vanishes everywhere),
that, if $M$ is such a matrix (with identity diagonal) then, all its powers (positive, negative and fractional) are 
substitution matrices and form a one-parameter group of substitutions, thus coming from a vector field on the line which can 
be computed.\\ 
But no nice combinatorial principle seems to emerge.\\ 
For example, beginning with the Stirling substitution 
$z\mapsto e^z-1$, we know that there is a unique one-parameter 
group of substitutions $s_\la(z)$ such that, for $\la$ integer, 
one has the value ($s_2(z) \lra \rm{partition of partitions}$)

\begin{equation}
s_2(z)=e^{e^{z}-1}-1\ ;\ s_3(z)=e^{e^{e^{z}-1}-1}-1\ ;\ s_{-1}(z)=ln(1+z)
\end{equation}
but we have no nice description of this group nor of the vector field generating it.

\subsection{A product formula}
The Hadamard product of two sequences $(a_n)_{n\geq 0};\ (b_n)_{n\geq 0}$ is given by the pointwise product $(a_nb_n)_{n\geq 0}$. 
We can at once transfer this law on EGFs by

\begin{equation}
\Big(\sum_{n\geq 0}a_n \frac{x^n}{n!}\Big)\ \odot_{exp}\ 
\Big(\sum_{n\geq 0}b_n \frac{x^n}{n!}\Big)
:=\sum_{n\geq 0}a_nb_n \frac{x^n}{n!}
\end{equation}
In the following, we will omit the subscript (in $\odot_{exp}$) as this will be the only kind of Hadamard product under consideration. 

But, it is not difficult to check that the family 
\begin{equation}
\left(\frac{(y\frac{\partial}{\partial x})^n}{n!}\ \frac{x^m}{m!}\right)_{n,m\in \N}
\end{equation}
is summable in $\C[[x,y]]$ (the space of formal power series in $x$ and $y$) as we have 
\begin{equation}
\frac{(y\frac{\partial}{\partial x})^n}{n!}\ \frac{x^m}{m!}=
\left\{ 
\begin{array}{cl}
0 				&\mathrm{if}\ n>m\\
\frac{y^nx^{m-n}}{n!(m-n)!}	&\mathrm{otherwise}
\end{array}
\right.
\end{equation}
and therefore, for $F(x)=\sum_{n\geq 0}a_n\frac{x^n}{n!}$ and $G(x)=\sum_{n\geq 0}b_n\frac{x^n}{n!}$ one gets the {\it product formula}

\begin{equation}
(F\odot G)(x):=F(y\frac{\partial}{\partial x})G(x)\big|_{x=0}=\sum_{n\geq 0}a_nb_n \frac{y^n}{n!}
\end{equation}
With this product, the set of series forms a commutative 
associative algebra with unit, which is actually the product algebra $\C^\N$.

\subsection{The double exponential formula}

The case $F(0)=G(0)=1$ will be of special interest in our study. Every series with constant term $1$ can be represented by 
an exponential $exp(\sum_{n\geq 1}L_n \frac{x^n}{n!})$ which can be expanded using Bell polynomials and Fa\`a di Bruno coefficients. 
Let us now recall some facts about these combinatorial notions.\\
We still consider the alphabet $\L=\{L_1,L_2,\cdots\}=\{L_i\}_{i\geq 1}$, then the complete Bell polynomials \cite{Co1} are defined by
\begin{equation}\label{bellpol}
exp(\sum_{m\geq 1} L_m \frac{x^m}{m!})=\sum_{n\geq 0} Y_n(\L) \frac{x^n}{n!}
\end{equation}
We will denote alternatively $Y_n(L_1,\cdots L_n)$ for $Y_n(\L)$ as this polynomial is independent from the subalphabet $(L_m)_{m>n}$. We 
know \cite{Co1} that 
\begin{equation}\label{bellpoldev}
Y_n(\L)=Y_n(L_1,\cdots L_n)=\sum_{||\al||=n}((\al)) \L^\al=\sum_{||\al||=n}((\al)) L_1^{\al_1}L_2^{\al_2}\cdots L_n^{\al_n}
\end{equation}
where $\al=(\al_1,\al_2,\cdots \al_n)$ is an integral vector, $||\al||:=\sum_{j=1}^m j\al_j$, $\L^\al=L_1^{\al_1}L_2^{\al_2}\cdots L_n^{\al_n}$ 
is the multiindex standard notation and 
\begin{equation}
((\al))=\frac{||\al||!}{(1!)^{\al_1}(2!)^{\al_2}\cdots (n!)^{\al_n}(\al_1)!\cdots (\al_n)!}
\end{equation}
is the Fa\`a di Bruno coefficient \cite{Co2,JR} which will be interpreted, in the next section, as enumerating structures called {\it set partitions}.\\
Combining \mref{bellpol} and \mref{bellpoldev} one gets

\begin{equation}\label{def}\!\!\!\!\!\!\!\!\!\!\!\!\!\!\!\!\!\!\!\!\!\!\!\!\!\!\!\!\!\!\!\!
exp\Bigg(\sum_{m\geq 1} L_m \frac{(y\frac{\partial}{\partial x})^m}{m!}\Bigg)exp\Bigg(\sum_{n\geq 1} V_n \frac{x^n}{n!}\Bigg)\bigg|_{x=0}=
\sum_{k\geq 0} \frac{y^k}{k!}\Bigg(\sum_{||\al||=||\be||} ((\al))((\be))\L^\al\V^\be\Bigg)
\end{equation}
Formula \mref{def} will be called in the sequel the {\it double exponential formula}. 

\subsection{Monomial expansion of the double exponential formula}\label{monexp}

In this paragraph, we will use unordered and ordered set partitions. By an unordered partition $P$ of the set $X$ we mean a finite subset 
$P\subset(\frak{P}(X)-\{\emptyset\})$ ($\frak{P}(X)$ is the set of all subsets of $X$ \cite{Co1}) such that 
\begin{equation}
\bigcup_{Y\in P}Y=X \rm{ and }(Y_1,Y_2\in P,\ Y_1\not=Y_2 \Longrightarrow Y_1\cap Y_2=\emptyset)
\end{equation}
this explains why {\it without any convention} the classical Stirling number of second the kind $S(0,0)$ equals $1$. The elements of $P$ are called 
{\it blocks}.\\ 
Following Comtet (\cite{Co2} p 39), we will say that a partition $P$ is of type $\al=(\al_1,\al_2,\cdots ,\al_m)$ iff there is no $j$-block 
for $j>m$ and $\al_j$ $j$-block(s) for each $j\leq m$. This implies in particular that the set $X$ is of cardinality $||\al||:=\sum_{j=1}^m j\al_j$.

Here one can see easily that the number of blocks of a partition of type $\al$ is $|\al|=\sum_{j=1}^m \al_j$.
An ordered partition of type $\al$ of the set $X$ is just a partition in which the blocks are labelled from $1$ to $|\al|$. \\
In other words, one could say that an ordered partition is a {\it list} of subsets and an unordered partition is a {\it set} of subsets.\\ 
To every ordered partition $P=(B_1,B_2,\cdots ,B_{|\al|})$ corresponds an unordered one\\ 
$\Phi_p(P)=\{B_1,B_2,\cdots ,B_{|\al|}\}$  
where $\Phi_p$ is the ``forgetful'' function which forgets the order. Now to a pair $(P^{(1)},P^{(2)})$ of ordered partitions of the same set (call it $X$)
\begin{equation}
P^{(1)}=(B^{(1)}_1,B^{(1)}_2,\cdots ,B^{(1)}_{k_1})\hspace{2cm} P^{(2)}=(B^{(2)}_1,B^{(2)}_2,\cdots ,B^{(2)}_{k_2})
\end{equation}
one can associate the intersection matrix $IM_o(P^{(1)},P^{(2)})$ such that the entry of address $(i,j)$ is the number of elements of the intersection of the block $i$ of the 
first partition and the block $j$ of the second. For example with partitions of $X=\{1,2,3,4,5,6\}$, and specifying  
$$P^{(1)}=\big(\{1,2,5\},\{3,4,6\}\big)\hspace{2cm} P^{(2)}=\big(\{1,2\},\{3,4\},\{5,6\}\big),$$ 
one gets 
\begin{center}
\begin{tabular}{c|c|c|c|}
   		& $\{1,2\}$ 	& $\{3,4\}$ 	& $\{5,6\}$\\
\hline
$\{1,2,5\}$	&  2		& 0		& 1\\	
\hline
$\{3,4,6\}$	&  0		& 2		& 1\\	
\hline
\end{tabular}
\end{center}
and hence the matrix 
\begin{equation}
\left(\begin{array}{ccc}
2 & 0 & 1\\
0 & 2 & 1
\end{array}\right)
\end{equation}
Formally, $IM_o(P^{(1)},P^{(2)})$ is the matrix of size $k_1\times k_2$ such that 
\begin{equation}
IM_o(P^{(1)},P^{(2)})[i,j]=\rm{card}\big(B^{(1)}_i\cap B^{(2)}_j\big)
\end{equation}
The matrices obtained in such a way form the set of packed matrices defined in  \cite{DHT} as, indeed, one sees that every packed matrix 
can be obtained through the matching procedure illustrated above.\\
If we consider now a pair of unordered partitions $(Q^{(1)},Q^{(2)})$, we cannot associate to them a single matrix but rather a class of matrices obtained 
from the preimages of $(Q^{(1)},Q^{(2)})$ under $\Phi_p\times \Phi_p$. In a compact formulation, the set of matrices so obtained is
\begin{equation}
\Big\{IM_0(P^{(1)},P^{(2)})\Big\}_{\Phi_p(P^{(1)})=Q^{(1)};\ \Phi_p(P^{(2)})=Q^{(2)}  }
\end{equation}
For example, with 
\begin{equation}
(Q^{(1)},Q^{(2)})=\big(\left\{\{1,2,5\},\{3,4,6\}\right\},\left\{\{1,2\},\{3,4\},\{5,6\}\right\}\big)
\end{equation}
one gets the 12 preimages $(P^{(1)}_i,P^{(2)}_j)$, where $P^{(1)}_i$ are among the two preimages of $Q^{(1)}$ and 
$P^{(2)}_j$ are among the 6 preimages of $Q^{(2)}$. Explicitely 
\begin{eqnarray*}
P^{(1)}_1=(\{1,2,5\},\{3,4,6\}) && P^{(1)}_2=(\{3,4,6\},\{1,2,5\})
\end{eqnarray*}
are the preimages of $Q^{(1)}$ and
\begin{eqnarray*}
P^{(2)}_1=(\{1,2\},\{3,4\},\{5,6\}) && P^{(2)}_2=(\{1,2\},\{5,6\},\{3,4\})\\
P^{(2)}_3=(\{3,4\},\{1,2\},\{5,6\}) &&  P^{(2)}_4=(\{3,4\},\{5,6\},\{1,2\})\\
P^{(2)}_5=(\{5,6\},\{1,2\},\{3,4\}) && P^{(2)}_6=(\{5,6\},\{3,4\},\{1,2\})
\end{eqnarray*}
are the preimages of $Q^{(2)}$.\\
The set of matrices so obtained reads
\begin{eqnarray}
IM_u(M)&=&\bigg\{
\left(\begin{array}{ccc}2 & 0 & 1\\0 & 2 & 1\end{array}\right),
\left(\begin{array}{ccc}2 & 1 & 0\\0 & 1& 2\end{array}\right),
\left(\begin{array}{ccc}1 & 2 & 0\\1 & 0 & 2\end{array}\right),\cr
&&
\left(\begin{array}{ccc}0 & 2 & 1\\2 & 0 & 1\end{array}\right),
\left(\begin{array}{ccc}0 & 1 & 2\\2 & 1 & 0\end{array}\right),
\left(\begin{array}{ccc}1 & 0 & 2\\1 & 2 & 0\end{array}\right)
\bigg\}.
\end{eqnarray}
This is the orbit of one of them under permutation of lines and columns.\\
The correspondence which, to a pair of unordered partitions, associates a class of matrices (under permutations of lines and columns) will be 
denoted $IM_u$.

\ss
Thus, one gets a commutative diagram of mappings
\begin{equation}
\begin{CD}
\rm{Pairs of ordered partitions} @> \Phi_p\times \Phi_p >> \rm{Pairs of unordered partitions} \\
@V{IM_o}VV						@VV{IM_u}V\\
\rm{Packed matrices}                 @> Class >> \rm{Classes of packed matrices}\\
@V{Dg_o}VV						@VV{Dg_u}V\\
\rm{Labelled diagrams}                 @> \Phi_d>> \rm{Diagrams}\\
\end{CD}
\end{equation}

The scheme presented above shows how to associate to a pair of ordered (resp. unordered) set partitions, a packed matrix (resp. a class of packed 
matrices). The packed matrices can be alternatively represented by {\it labelled diagrams} which are bipartite multigraphs built from two sets of 
vertices being a column of white spots (WS) and column of black spots (BS) as shown below.

\vspace{5mm}
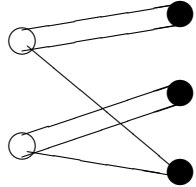
\begin{center}\label{fig1}
\setlength{\unitlength}{0.7mm}
\begin{picture}(40,40)
\put(5,10){\circle{5}}
\put(5,30){\circle{5}}

\put(35,5){\circle*{5}}
\put(35,20){\circle*{5}}
\put(35,35){\circle*{5}}

\put(5,32){\line(6,1){30.41}}
\put(5,28){\line(6,1){30.41}}
\put(5,12){\line(3,1){30.41}}
\put(5,08){\line(3,1){30.41}}

\put(6,29){\line(6,-5){30}}
\put(6,9){\line(6,-1){27}}

\end{picture}

\bs
Labelled diagram of the matrix  $\left(\begin{array}{ccc}2 & 0 & 1\\0 & 2 & 1\end{array}\right)$
\end{center}
\vspace{1cm}

Let us explain how to associate to a (drawn) diagram a packed matrix. 
The white (resp. black) spots are labelled from $1$ to $r$ (resp. $1$ to $c$) from top to bottom and the number of lines 
from the $i$-th white spot to the $j$-th black spot is exactly the entry $a_{ij}$ of the matrix. Conversely, a packed matrix of dimension $r\times c$ 
being given, one draws $r$ white spots (resp. $c$ black spots) and (with the labelling as above) join the $i$-th white spot to the $j$-th black 
by $a_{ij}$ lines. This gives exactly the one-to-one correspondence between (drawn) diagrams and packed matrices. 

In the sequel, we set $Diag_u:=Dg_u\circ IM_u$ and $Diag_o:=Dg_o\circ IM_o$ for the mappings which associate diagrams to pairs of partitions.
Now, the multiplicity of a diagram $\D$ is the number of pairs $(P^{(1)},P^{(2)})$ of unordered partitions such that $Dg_u(IM_u(P^{(1)},P^{(2)}))=\D$.\\
Let us call bitype of a diagram $\D$ the pair $(\al(P^{(1)},\al(P^{(2)})$ where $Dg_u(IM_u(P^{(1)},P^{(2)}))=\D$ (remark that it does not depend on the choosen premiage inside the formula) 
and we will refer it as the bitype $(\al(\D),\be(\D))$) of $\D$. In a similar way $\al(\D)$ (resp. $\be(\D)$) will be called the left (resp. the right) type of $\D$.\\
The product formula now reads
\begin{eqnarray}\label{prodform}\!\!\!\!\!\!\!\!\!\!\!\!\!\!\!\!\!\!\!\!\!\!\!\!
exp\Big(\sum_{m\geq 1}\frac{L_m}{m!}\left(y \frac{\partial}{\partial x}\right)^m\Big) exp\Big(\sum_{n\geq 1}\frac{V_n}{n!}x^n\Big)\Bigg|_{x=0}=
\cr\!\!\!\!\!\!\!\!\!\!\!\!\!\!\!\!\!\!\!\!\!\!\!\!
\sum_{n\geq 0} \frac{y^n}{n!} \Big(\!\sum_{\D\rm{ diagram}\atop |\D|=n}mult(\D) \mathbb{L}^{\al(\D)}\mathbb{V}^{\be(\D)}\!\Big)\!=\!
\!\sum_{\D\ \rm{diagram}}\!\!\frac{mult(\D)}{|\D|!}m(\D,\L,\V,y)
\end{eqnarray}
with 
\begin{equation}\label{m}
m(\D,\L,\V,y):=\L^{\al(\D)}\V^{\be(\D)} y^{|\D|}.
\end{equation}

\section{Diagrammatic expansion of the double exponential formula}\label{diagexp}

The main interest of the expansion \mref{prodform} is that we can impose (at least) two types of rules on the diagrams 
\begin{itemize}
\item on the diagrams themselves (selection rules) : on the outgoing degrees, ingoing degrees, total or partial weights (the graph is 
supposed oriented from white to black spots)
\item on the set of diagrams (composition and decomposition rules): product and coproduct on the space of diagrams.
\end{itemize}

We have already such a structure on the space of monomials (i.e. the polynomials). \\
The (usual) product of polynomials is well known and amounts to the addition of the multidegrees. The (usual) coproduct is given by the 
substitution of a ``doubled'' variable to each variable \cite{Bour1,JR}. For example, with $P=x^2y^3$, we first form $(x_1+x_2)^2(y_1+y_2)^3$, expand and then 
separate (on the left) the ``$1$'' labelled variables and (on the right) the ``$2$'' labelled. As
\begin{eqnarray}
P=x_1^2y_1^3+3x_1^2y_1^2y_2+3x_1^2y_1y_2^2+x_1^2y_2^3+2x_1y_1^3x_2+
6x_1y_
1^2x_2y_2+
\cr 6x_1y_1x_2y_2^2+
2x_1x_2y_2^3+y_1^3x_2^2+3y_1^2x_2^2y_2+
3y_1x_2^2y_2^2+x_2^2y_2^3
\end{eqnarray}

one gets, with $\Delta$ the coproduct operator, 

\begin{eqnarray}
\Delta(P)&=&x^2y^3\otimes 1+3x^2y^2\otimes y+3x^2y\otimes y^2+x^2\otimes y^3+
\cr
&&
2xy^3\otimes x+6xy^2\otimes xy+6xy\otimes xy^2+ 2x\otimes xy^3+
\cr
&&
y^3\otimes x^2+3y^2\otimes x^2y+ 3y\otimes x^2y^2+1 \otimes x^2y^3.
\end{eqnarray}

The space of polynomials with product and coproduct (and other items like neutrals, co-neutrals and antipode,  
which will be made more precise in the next paragraph) is endowed with the structure of a Hopf algebra.\\
The last consideration suggests the following question: 

\ss
{\it Is it possible to structure the (spaces of) diagrams into a Hopf algebra ? Is it possible that this structure be compatible, in some sense,
  with the mapping $(\D,\L,\V,y)\mapsto m(\D,\L,\V,y)$ ?
}

\ss
Answer is yes. To establish it, we have to proceed in three steps. 

\begin{itemize}
\item First Step : Define the space(s)
\item Second Step : Define a product
\item Third Step : Define a coproduct
\end{itemize}

\subsection{Algebra structure}

{\sc First step}\pointir Let $\Diag_\C$ (resp. $LDiag_\C$) be the $\C$-vector space freely generated by the diagrams (resp. labelled diagrams) i.e. 

\begin{equation}
\Diag_\C:=\bigoplus_{d\ \rm{diagram}}\C \hspace{2cm} \LDiag_\C:=\bigoplus_{d\ \rm{labelled diagram}} \C d 
\end{equation}
at this stage, we have a linear mapping (linear arrow) $\LDiag_\C\mapsto \Diag_\C$ provided by the linear extension of $\Phi_d$ and an arrow 
(linear, by construction) $m(.,\L,\V,z)\!:\Diag_\C\mapsto \C[\L\cup\V\cup \{z\}]$ provided by the linear extension of $m(.,\L,\V,z)$. 

\bs
{\sc Second step}\pointir We remark that, if

\begin{equation}\label{juxt}
d_1\star d_2=\fbox{$\begin{array}{c} d_1\\ d_2 \end{array}$}
\end{equation}

denotes the superposition of the diagrams, then 

\begin{equation}\label{prodmon}
m(d_1\star d_2,\L,\V,z)=m(d_1,\L,\V,z)m(d_2,\L,\V,z).
\end{equation}

The law \mref{juxt} makes sense as well for labelled and unlabelled diagrams. In the first case, it amounts to computing the blockdiagonal 
product of packed matrices. Indeed, for $M_1,\ M_2$ being packed matrices, one has 
\begin{equation}
Dg_o\left(\left(\begin{array}{cc}M_1 & 0\cr 0 & M_2\end{array}\right)\right)=Dg_o(M_1)\star Dg_o(M_2).
\end{equation}

This product yields the product of monomials in the following way. From $\D$ a diagram and all the other parameters fixed, with 
the setting of \mref{m}, we get a polynomial. 

The product \mref{juxt} is associative with unit (the empty diagram), it is compatible with the arrow $\Phi_d$ and so defines the product on 
$\Diag$ which, in turn is compatible with the product of monomials.

\begin{equation}
\begin{CD}
\rm{Labelled diagrams}^2 @> \Phi_d\times \Phi_d >> \rm{Diagrams}^2 @> m(?,\L,\V,z)\times m(?,\L,\V,z) >> \rm{Monomials}^2\\
@ V\rm{product}VV				 @ V\rm{product}VV		@ V\rm{product}VV\\
\rm{Labelled diagrams} @> \Phi_d >> \rm{Diagrams} @> m(?,\L,\V,z) >> \rm{Monomials}\\
\end{CD}
\end{equation}
 
\begin{rem} One sees easily that the labelled diagram (resp. diagrams) form monoids thus the spaces 
 $\LDiag_\C$ and $\Diag_\C$ are algebras of these monoids \cite{BDK,BR}.
\end{rem}
\subsection{Admissible coproducts}

For the coproduct on $\LDiag$, we have several possibilities:

\begin{enumerate}
\item split with respect to the white spots (two ways : by intervals and by subsets) 
\item split with respect to the black spots (two ways : by intervals and by subsets) 
\item split with respect to the edges
\end{enumerate}

The discussion goes as follows:\\ 
i) (3) does not give a nice identity with the monomials (when applying $d\mapsto m(d,?,?,?)$) nor do (2) and (3) by intervals.\\ 
ii) (2) and (3) are essentially the same (because of the WS $\lra$ BS symmetry).\\
In fact (2) and (3) by subsets give a good representation and, moreover, they are appropriate  for several physics models. 

In the next section, we develop the possibility (1) and (2) by subsets.

\section{Hopf algebra structures associated with $\Delta_{WS}$ and $\Delta_{BS}$}\label{hopf}

\subsection{The philosophy of bi- and Hopf algebras thru representation theory}

Let $\A$ is a $k$-algebra ($k$ is a field as $\R$ or $\C$). In this paragraph, we consider associative algebras with unit (AAU). 
A representation of $\A$ is here a pair $(V,\rho_V)$ where $V$ is a $k$-vector space and $\rho_V: \A\mapsto End_k(V)$ a morphism of $k$-algebras 
(AAU).\\
One can make operations with representations as direct sums and quotient of a representation by a sub-representation (a sub-representation is 
a subspace which is closed under the action of $\A$).  
In general, one does not know how to endow the tensor product (of two representations) and the dual (of a representation) with the structure of 
representation.\\
It is however classical in two cases: groups and Lie algebras.\\
If $G$ is a group, a representation of $G$ is a pair $(V,\rho_V)$ where $V$ is a $k$-vector space and $\rho_V: G\mapsto Aut_k(V)$ a morphism 
of groups. If $G$ is a Lie algebra, a representation of $G$ is a pair $(V,\rho_V)$ where $V$ is a $k$-vector space and $\rho_V: G\mapsto End_k(V)$ 
a morphism of Lie algebras (i.e. $\rho_V([u,v])=\rho_V(u)\rho_V(v)-\rho_V(v)\rho_V(u)$). These two cases enter the scheme of (AAU) as a representation 
of a group can be extended uniquely as a representation of its algebra $kG$ and a representation of a Lie algebra as a representation of $\U_k(G)$, its 
envelopping algebra. These two constructions ($kG$ and $\U_k(G)$) are (AAU).\\
For the sake of readibility let us denote in all cases $\rho_V(g)(u)$ by $g.u$ ($g\in G$ and $u\in V$).\\
If $G$ is a group and $V,W$ two representations, we construct a representation of $G$ on $V\otimes W$ by 
\begin{equation}
g.(u\otimes v)=g.u\otimes g.v
\end{equation}
If $G$ is a Lie algebra and $V,W$ two representations, we construct a representation of $G$ on $V\otimes W$ by 
\begin{equation}
g.(u\otimes v)=g.u\otimes v + u\otimes g.v
\end{equation}
This can be rephrased in saying that the action of $g$ in the first case (group) is $g\otimes g$ and in the second (Lie algebra) $g\otimes 1+1\otimes g$ 
($1$ is here for the appropriate identity mapping). In the two cases, it amounts to give a linear mapping $\Delta: \A\mapsto \A\otimes \A$ which will be 
called a coproduct.\\ 
One can show \cite{CP} that, if we want that this new operation enjoy ``nice'' properties (associativity of the tensor product etc...), 
one has to suppose that this coproduct is a morphism of (AAU) ($\A\otimes \A$ has received the structure of - non twisted - tensor product of algebras), 
that it is coassociative with a counit \cite{CP}. Let us make these requirements more precise.\\
The first says that for all $x,y\in \A$ one has $\Delta(xy)=\Delta(x)\Delta(y)$, the second that the two compositions
\begin{equation}
\A \stackrel{\Delta}{\longrightarrow} \A\otimes \A \stackrel{\Delta\otimes 1_\A}{\longrightarrow}  \A\otimes \A \otimes \A \rm{ and }
\A \stackrel{\Delta}{\longrightarrow} \A\otimes \A \stackrel{1_\A\otimes \Delta}{\longrightarrow}  \A\otimes \A \otimes \A 
\end{equation}
are equal, the third says that there is a mapping (linear form) $\ep: \A \mapsto k$ such that the compositions
\begin{equation}
\A \stackrel{\Delta}{\longrightarrow} \A\otimes \A \stackrel{\Delta\otimes \ep}{\longrightarrow}  \A\otimes k \stackrel{\rm{nat}}{\longrightarrow} \A
\rm{ and }
\A \stackrel{\Delta}{\longrightarrow} \A\otimes \A \stackrel{\ep\otimes \Delta}{\longrightarrow}  k\otimes \A \stackrel{\rm{nat}}{\longrightarrow} \A
\end{equation}
(where $\rm{nat}$ is for the natural mappings) are equal to the identity $1_\A$.\\
An algebra (AAU) together with a coproduct $\Delta$ and a counity $\ep$ which fulfills the three requirements above is called a {\bf bialgebra}.

\bs
If, moreover one wants to have a nice dualization of the representations (i.e. nice structures for the duals $V^*=Hom(V,k)$), it should exist an 
element of $Hom(\A,\A)$ such that the compositions
\begin{equation}
\A \stackrel{\Delta}{\longrightarrow} \A\otimes \A \stackrel{\al\otimes 1_\A}{\longrightarrow}  \A\otimes \A 
\stackrel{\mu}{\longrightarrow} \A\rm{ and }
\A \stackrel{\Delta}{\longrightarrow} \A\otimes \A \stackrel{1_\A \otimes \al}{\longrightarrow}  \A\otimes \A 
\stackrel{\mu}{\longrightarrow} \A
\end{equation}
are equal to $e_A\ep$ (where $e_A$ denotes the unit of $A$). When a bialgebra possesses such an element (unique), it  is called 
the {\bf antipode} and the bialgebra a {\bf Hopf algebra}. For more details and connections to physics, one can consult \cite{CP}. 

\ss
One can prove that the bialgebras constructed below possess an antipode and then are Hopf algebras.

\subsection{Bialgebra structures on $LDiag$ and $Diag$}

The space spanned by the packed matrices has already received a structure of Hopf algebra, the algebra $\MQSym$ \cite{DHT}. We briefly review 
the structure of this Hopf algebra.\\
We describe in details $\DWS$ as the other coproduct is actually got by the same process but applied on the columns instead of the lines. 
Let $M$ be a packed matrix of dimensions $k_1\times k_2$ for every subset $X\in [1..k_1]$ we consider the matrix $\pi_X(M):=pack(M[X,[1..k_2]])$, 
the restriction to the lines of $X$ and then packed (with this restriction to the lines, we only need to perform a horizontal packing). Thus, the 
coproduct $\DWS$ reads
\begin{equation}
\DWS(M)=\sum_{X+Y=[1..k_1]}\pi_X(M)\otimes \pi_Y(M)
\end{equation}
To avoid confusion we will call the supporting space $\HWS$ ($=\MQSym$). We keep the (total) grading of $\MQSym$ by the total weight (i.e. the 
sum of the coefficients) of the matrices. The packed matrices are a linear basis of $\HWS=\MQSym$, thus every element expresses uniquely
\begin{equation}
x=\sum_{M \rm{ packed}} \lambda_M(x) M
\end{equation}
The coproduct above is cocommutative and with counit $\lambda_{\emptyset}$ where $\emptyset$ is the void matrix corresponding to 
the void diagram. This particular matrix will be denoted $\1H$.\\
For example, with the packed matrix above one has 

\begin{eqnarray*}
\DWS(\left(\!\!\begin{array}{cc} 2 & 0\\0 & 2\\1 & 1\end{array}\!\!\right))&=&\left(\!\!\begin{array}{cc} 2 & 0\\0 & 2\\1 & 1\end{array}\!\!\right)\otimes \1H + 
\left(\!\!\begin{array}{c} 2\end{array}\!\!\right)\otimes 
\left(\!\!\begin{array}{cc} 0 & 2\\1 & 1\end{array}\!\!\right) + 
\left(\!\!\begin{array}{c} 2\end{array}\!\!\right)\otimes 
\left(\!\!\begin{array}{cc} 2 & 0\\1 & 1\end{array}\!\!\right) +
\cr
&&
\left(\!\!\begin{array}{cc} 1 & 1\end{array}\!\!\right)
\otimes \left(\!\!\begin{array}{cc} 2 & 0 \\ 0 & 2\end{array}\!\!\right) +
\left(\!\!\begin{array}{cc} 2 & 0 \\ 0 & 2\end{array}\!\!\right)\otimes \left(\!\!\begin{array}{cc} 1 & 1\end{array}\!\!\right) +
\cr&& 
\left(\!\!\begin{array}{cc} 2 & 0 \\ 1 & 1\end{array}\!\!\right)\otimes \left(\!\!\begin{array}{c} 2\end{array}\!\!\right) +
\left(\!\!\begin{array}{cc} 0 & 2 \\ 1 & 1\end{array}\!\!\right)\otimes \left(\!\!\begin{array}{c} 2\end{array}\!\!\right) + 
\1H \otimes \left(\begin{array}{cc} 2 & 0\\0 & 2\\1 & 1\end{array}\!\!\right)\end{eqnarray*}

This coproduct is compatible with the usual coproduct on the monomials for the constant alphabet $\V=\mathbf{1}_\N$ defined by $V_n=1$ for all 
$n\geq 0$. Then, using Sweedler's notation, for this particular $\V$, if $\DWS(d)=\sum d_{(1)}\otimes d_{(2)}$, one has
\begin{equation}
m(d,\L'+\L'',\mathbf{1}_\N,z)=\sum m(d_{(1)},\L',\mathbf{1}_\N,z)m(d_{(2)},\L'',\mathbf{1}_\N,z)
\end{equation}

Thus, one sees easily that, with this structure (product with unit, coproduct 
and the counit), $\LDiag_\C$ is a bialgebra graded in finite dimensions and then a Hopf algebra.\\ 
The arrow $\LDiag_\C\mapsto \Diag_\C$ endows $\Diag_\C$ with a structure of Hopf algebra. 

\section{Conclusion}

The structure of the Hopf algebras $\LDiag_\C, \Diag_\C$, by a theorem of Cartier, Milnor and Moore \cite{Ca, MM}, is that of envelopping algebras of their 
primitive elements ($\Diag_\C$, being commutative, is thus an algebra of polynomials).\\
Moreover, it appears that the structure described above is the starting point for a series of connections with mathematical and physical Hopf algebras. 
The coproduct $\Delta_{BS}$ is the cristallisation  ($q=1$) of a one-parameter deformation of coproducts (all coassociative) on 
$\LDiag_\C\simeq {\bf MQSym}$, the other end ($q=0$) being an infinitesimal coproduct isomorphic to $\Delta_{\bf MQSym}$. Recently, ${\bf FQSym}$ 
(a subalgebra of $\mathbf{MQSym}$) has been established by Foissy \cite{Fo1} as a case in a family of Hopf algebras of decorated planar trees 
which is strongly related to other Hopf algebras like Connes-Kreimer's and Connes-Moscovici's \cite{Fo1,Fo2}.

\end{document}